\documentclass[journal]{IEEEtran}
\ifCLASSINFOpdf
   \usepackage[pdftex]{graphicx}
   \usepackage{gensymb}
\else
\fi
\begin{document}
%
\title{SAW synthesis with IDTs array and the inverse filter: toward a versatile SAW toolbox for microfluidics and biological applications.}
%
%
%

\author{Antoine Riaud, Michael Baudoin, Jean-Louis Thomas, Olivier Bou Matar
\thanks{Manuscript received December 25, 2015; revised XXX; accepted XXX. This work was supported by Agence Nationale de la Recherche ANR-12-BS09-0021-01 and ANR-12-BS09-0021-02 and Region Nord Pas de Calais. }
\thanks{The authors are with Institut d'Electronique, de Micro\'{e}lectronique et de Nanotechnologie (IEMN), LIA LICS, Universit\'{e} de Lille and EC Lille, UMR CNRS 8520, 59652 Villeneuve d'Ascq, France and  Sorbonne Universit\'{e}s, UPMC Univ Paris 06, CNRS UMR 7588, Institut des NanoSciences de Paris (INSP), F-75005 Paris, France. Corresponding author : Michael Baudoin (email : michael.baudoin@univ-lille1.fr)}.
}

\maketitle

\begin{abstract}
Surface acoustic waves (SAWs) are versatile tools to manipulate fluids at small scales for microfluidics and biological applications. A non-exhaustive list of operations that can be performed with SAW includes sessile droplet displacement, atomization, division and merging but also the actuation of fluids embedded in microchannels or the manipulation of suspended particles. However, each of these operations requires a specific design of the wave generation system, the so-called interdigitated transducers (IDTs). Depending on the application, it might indeed be necessary to generate focused or plane, propagating or standing, aligned or shifted waves. Furthermore, the possibilities offered by more complex wave-fields such as acoustical vortices for particle tweezing and liquid twisting cannot be explored with classical IDTs. In this paper, we show that the inverse filter technique coupled with an interdigitated transducers array (IDTA) enables to synthesize all classical wave-fields used in microfluidics and biological applications with a single multi-function platform. It also enables to generate swirling SAWs, whose potential for the on-chip synthesis of tailored acoustical vortices has been demonstrated lately. The possibilities offered by this platform is illustrated by performing successively many operations on sessile droplets with the same system.
\end{abstract}


%
\IEEEpeerreviewmaketitle

\section{Introduction}
%
%
%
%
\IEEEPARstart{S}{ince} the seminal work of Shiokawa in 1989 on the atomization of droplets \cite{ieee_shiokawa_1989}, the potential of SAWs for fluid actuation at microscale has been widely explored in the literature: Progressive waves synthesized by straight IDTs enable sessile droplets displacement \cite{abc_wixforth_2004, saa_renaudin_2006, pre_brunet_2010,apl_baudoin_2012} or fluid pumping in a micro-channel \cite{loc_girardo_2008,apl_cecchini_2008}. Acoustic field generated by pairs of IDTs can be combined to synthesize stationary waves and manipulate collectively particles \cite{jap_shilton_2008,loc_huang_2008,lc_huang_2009,mn_raghavan_2010,apl_tran_2012,pnas_ding_2012,ieee_guo_2014,apl_collins_2014}, or cells \cite{loc_shi_2009,loc_franke_2010,loc_hartmann_2014,apl_sivanatha_2014,pnas_ding_2014,pnas_li_2015,pnas_guo_2015}  in droplets or embedded in a microfluidic chamber. Focused waves synthesized by concentric IDTs are suitable for the fusion of droplets \cite{mn_babetta_2012} or for high power applications such as droplet atomization \cite{prl_tan_2009}. Finally, more complex operations such as mixing at low Reynolds number through chaotic advection or droplet division may require some fancy combination of plane waves with either rotating \cite{prl_frommelt_2008} or  shifted \cite{loc_collignon_2015} wavefronts.  So far, it has thus been demonstrated that SAWs enable to perform many basic operations required in microfluidics and biological applications \cite{rmp_friend_2011,arfm_yeo_2014}.

Nevertheless each of these operations requires a specific optimized design, which is not compatible with a multi-function actuation platform required for the development of many labs-on-chips. Moreover, classical IDTs do not allow the synthesis of swirling SAWs \cite{prap_riaud_2015} envisioned for on-chip synthesis of tailored acoustical vortices \cite{pre_riaud_2015} and consequently three-dimensional single particle manipulation \cite{prl_baresch_2015} and vorticity control \cite{pre_riaud_2014}. To overcome these shortcomings, we developed an array of 32 optimized IDTs driven by a programmable electronics that enables independent control of each transducer. To synthesize the targeted wave-field, the inverse filter technique \cite{jasa_tanter_2000} originally developed for bulk waves has been adapted for SAWs \cite{prap_riaud_2015}. In this paper, we show that it is possible to synthesize plane waves in different directions, focused waves with focal point located at arbitrary position and swirling SAWs with the same device.  The potential of this system is illustrated on sessile droplets by showing successively droplet controlled displacement, division, fusion and nebulization with the same platform.

\section{Methods}

\begin{figure}[htbp]
\begin{center}
\includegraphics[width=0.4\textwidth]{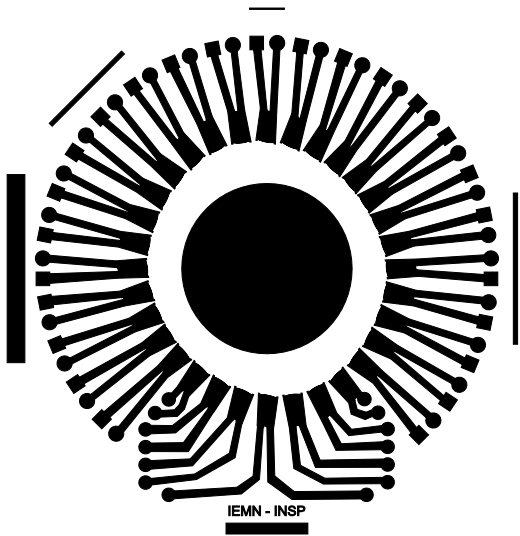}
\end{center}
\caption{Optimized array of 32 interdigitated unidirectional transducers used for the synthesis of various surface acoustic wave-fields (focused, plane, swirling) and the manipulation (displacement, division, fusion, atomization) of droplets. The central black zone is a gold layer used as a mirror for the measurement of the transducer response with the Michelson interferometer.}
\label{fig:array}
\end{figure}

Rayleigh surface acoustic waves (R-SAWs) are synthesized at the surface of a X-cut, 1.05 mm thick, Niobate Lithium (LiNb03) piezoelectric substrate by an array of 32 unidirectional interdigitated transducers (SPUDT IDTs) (see Fig. \ref{fig:array}).  The X-cut was chosen for its good electro-mechanical coefficients along Z $(K^2 = 5,9 \%)$ and Y $(K^2 = 3,1 \%)$ directions and its weaker anisotropy compared to the 128\degree \, Y-cut generally used for unidirectional IDTs. The IDTs array (IDTA) was fabricated using a lift-off process:  (i) the substrate is coated with a AZ15-10 photoresist sacrificial layer patterned with conventional photolithography technique, (ii) a titanium (Ti) layer of 20 nm and a gold (Au) layer of 200 nm are deposited by evaporation  (titanium is used for its good adherence on LiNbO3 and gold for its good electrical conductivity), (iii) the sacrificial layer is washed out by a developer.

The IDTA has been designed for a working frequency of $11.9$ MHz, with a wavelength adapted to each wave propagation direction of the anisotropic LiNbO3 crystal. Compared to previous work by the authors \cite{prap_riaud_2015}, the design has been further optimized by placing the IDTs on a slowness curve and slightly curving the IDTs to promote diffraction. These two modifications improve the illumination by each transducer of the central zone (radius of 5 mm) of the substrate that we call the "acoustical scene" and where microfluidic operations are performed. An optimal spatial coverage is indeed essential for the synthesis of a wide variety of acoustic wave-fields.

Each of the IDTs is excited independently with a dedicated programmable electronics that enables the synthesis of wave packets at carrying frequencies up to 12 MHz. Impedance matching for each transducers was achieved with external electronic components (inductances). Finally, the inverse filter method \cite{jasa_tanter_2000} was used to determine the optimal input signal for each IDT to synthesize a targeted wave-field. Indeed, the inverse filter is a very general method to synthesize a specific wave-field in a linear medium given a set of independent programmable sources. This process can be basically decomposed into three distinct stages. First, the signal emitted by each transducer (impulse response) is measured in the acoustics scene. In practice this response is recorded in a number of control points whose distance cannot exceed $\lambda / 2$ according to Nyquist-Shannon sampling principle. This allows to define, in the Fourier space, a transfer matrix $H_{ij}(\omega)$ (called the propagation operator)  between the Fourier transform of the entrance signal emitted by transducer $j$, $E_j(\omega)$, and the response signal at control point $i$, $S_i(\omega)$: $S_i = H_{ij} E_j$ (with Einstein notations). In the present experiments, the surface vibrations of the substrate at control point $i$ (typically of the order of a few nanometers) are measured by a home-made polarized Michelson interferometer whose principle is given in reference \cite{prap_riaud_2015}. Then, the targeted output signal $S$ is defined and the transfer matrix H is inverted to compute the optimal input signal $E = H^{-1} S$. Finally, the optimal signal is synthesized by each transducer.  If $e_j(t)$, $s_i(t)$ and $h_{ij}(t)$ are the inverse Fourier transform of $E_j(\omega)$, $S_i(\omega)$ and $H_{ij}(\omega)$, it is worth noting that the output time signal $s(t)$ is the convolution product of $h_{ij}(t)$ and $e_j(t)$: 
$$ s_i(t) = h_{ij}(t) * e_j(t).$$
While the inverse filter method is simple in principle, some complexity arises when implementing it. Indeed, the propagation operator is generally ill-conditioned since small errors in the measurements produce very large errors in the reconstructed results. Then, the number of control points is not necessarily the same as the number of sources (transducers) and thus the propagation operator is not necessarily a square matrix. So, the pseudo-inverse of the propagation operator is obtained through singular value decomposition. Finally, the inverse filter technique had been initially developed to generate acoustical wave-fields in 3D media. In this case, the target field is a surface and has a smaller dimensionality (2D) than the propagative medium (3D), whereas for SAWs the target field has the same dimensionality as the propagative medium (both 2D). So the control points are not independent and the wave-field must fulfill the dispersion relation. This requires some refinements in the method (see Ref. \cite{prap_riaud_2015}).

\begin{figure}[htbp]
\begin{center}
\includegraphics[width=0.4\textwidth]{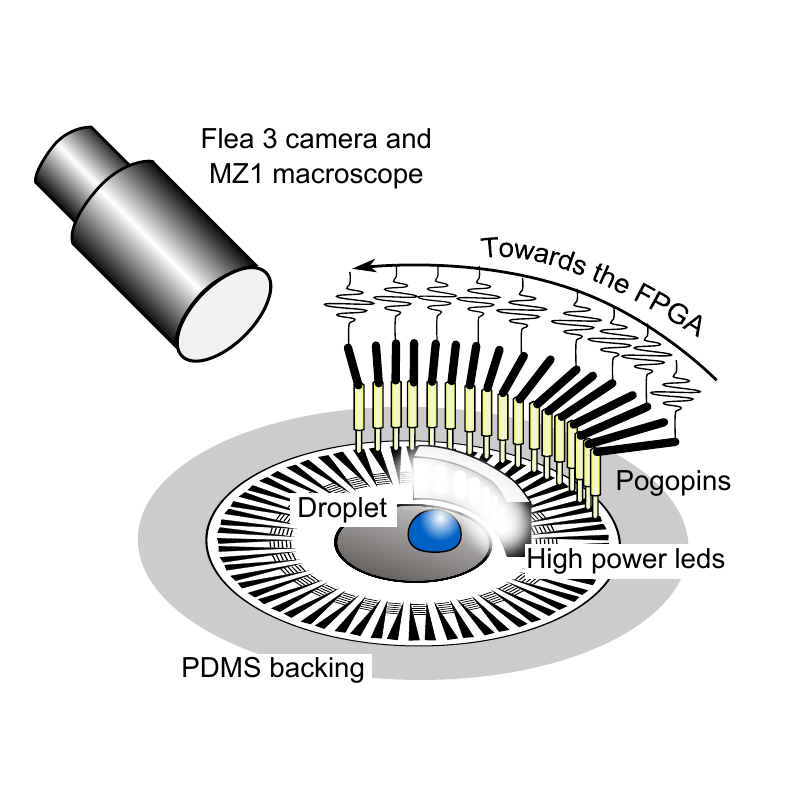}
\end{center}
\caption{Sketch of the experimental setup used for the synthesis of various wave-fields and the actuation of droplets.}
\label{fig:expe}
\end{figure}

To enable droplet manipulation, the central zone was treated with a hydrophobic self assembled monolayer. Otherwise, the droplet would spread on the gold layer at the center of the Niobate Lithium substrate since it is a perfectly wetting medium. Nevertheless, the adherence of alkane-thiol molecules on the central gold layer was not optimal leading to contact angle below $90 \degree$ ($\theta \approx 70 \degree$), large hysteresis and thus pining of the contact line. Despite these bad properties of the hydrophobic layer, we nevertheless succeeded to control the drop displacement with the IDTA as we will see in the next section. The droplet dynamics was recorded with a Flea 3 camera on a MZ1 macroscope at 75 frames per second. A sketch of the experimental setup is shown on Fig. \ref{fig:expe}.

\section{Results}

\subsection{Complex wave-fields synthesis}

To demonstrate the potential of the SAW toolbox for fluid samples actuation, we first investigated the possibilities offered by the IDTA and the inverse filter technique to synthesize the main surface acoustic wave-fields used in the literature: plane, focalized and swirling SAWs. 

\begin{figure}[htbp]
\begin{center}
\includegraphics[width=0.45\textwidth]{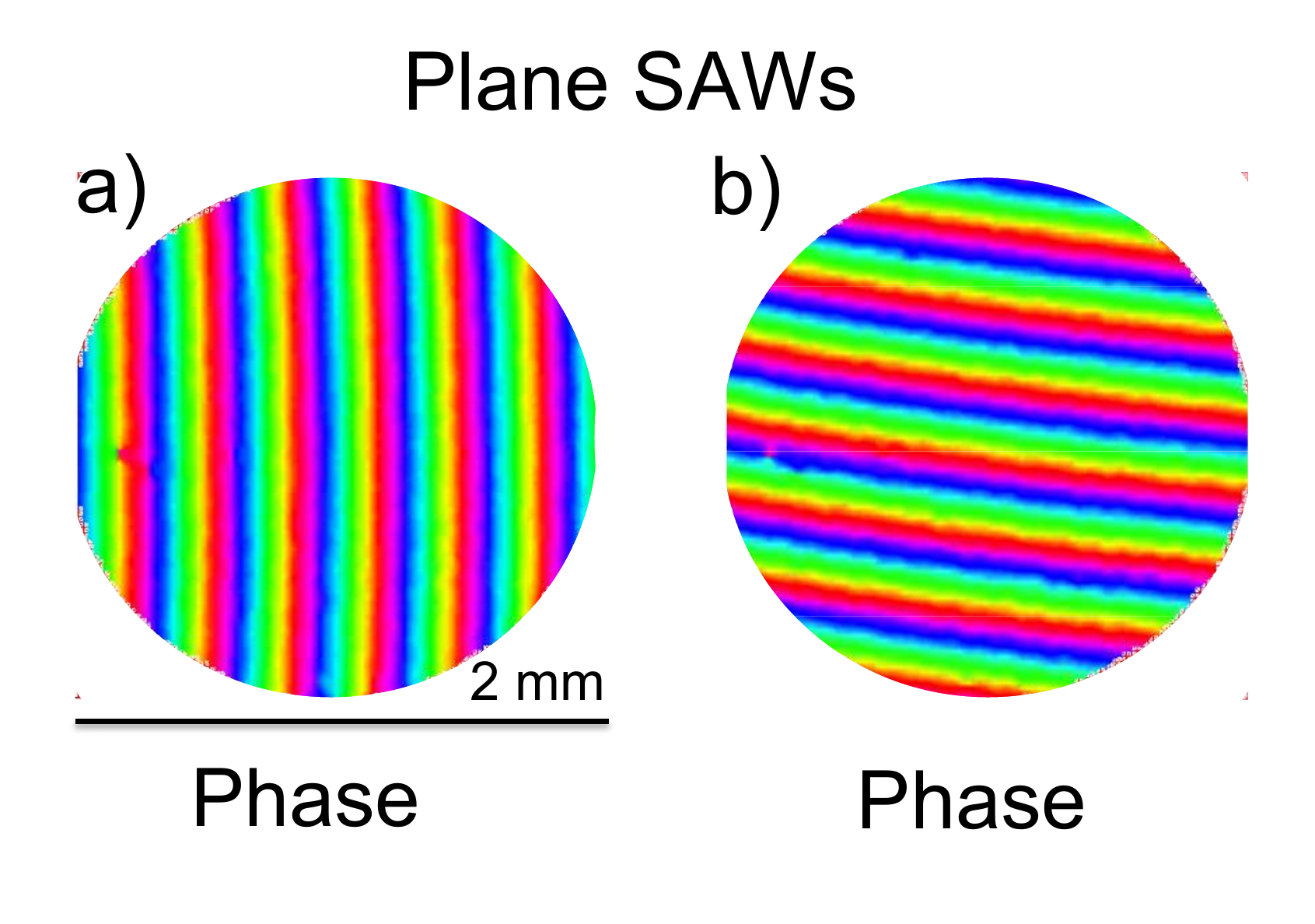}
\end{center}
\caption{Phase of plane progressive SAWs synthesized with the IDTA in two different directions (a and b). The maximum peak to peak amplitude in the two directions are respectively $6$ and $10$ nm.}
\label{fig:plane}
\end{figure}

While our setup is optimized for the synthesis of focalized waves and swirling SAWs (since the IDTs are disposed radially around the slowness curve), we have shown that it is possible to synthesize waves with plane wavefronts in the desired direction with peak to peak amplitudes larger than $5.5$ nm, that is to say well above the values classically used for droplets displacement \cite{pre_brunet_2010,apl_baudoin_2012}. We provide on Fig. \ref{fig:plane} two examples of plane progressive waves synthesized in different directions. These results show that despite the curvature of the IDTs, the wavefront aberration is weak in the acoustical scene and compensated by the inverse filter.  

\begin{figure}[htbp]
\begin{center}
\includegraphics[width=0.45\textwidth]{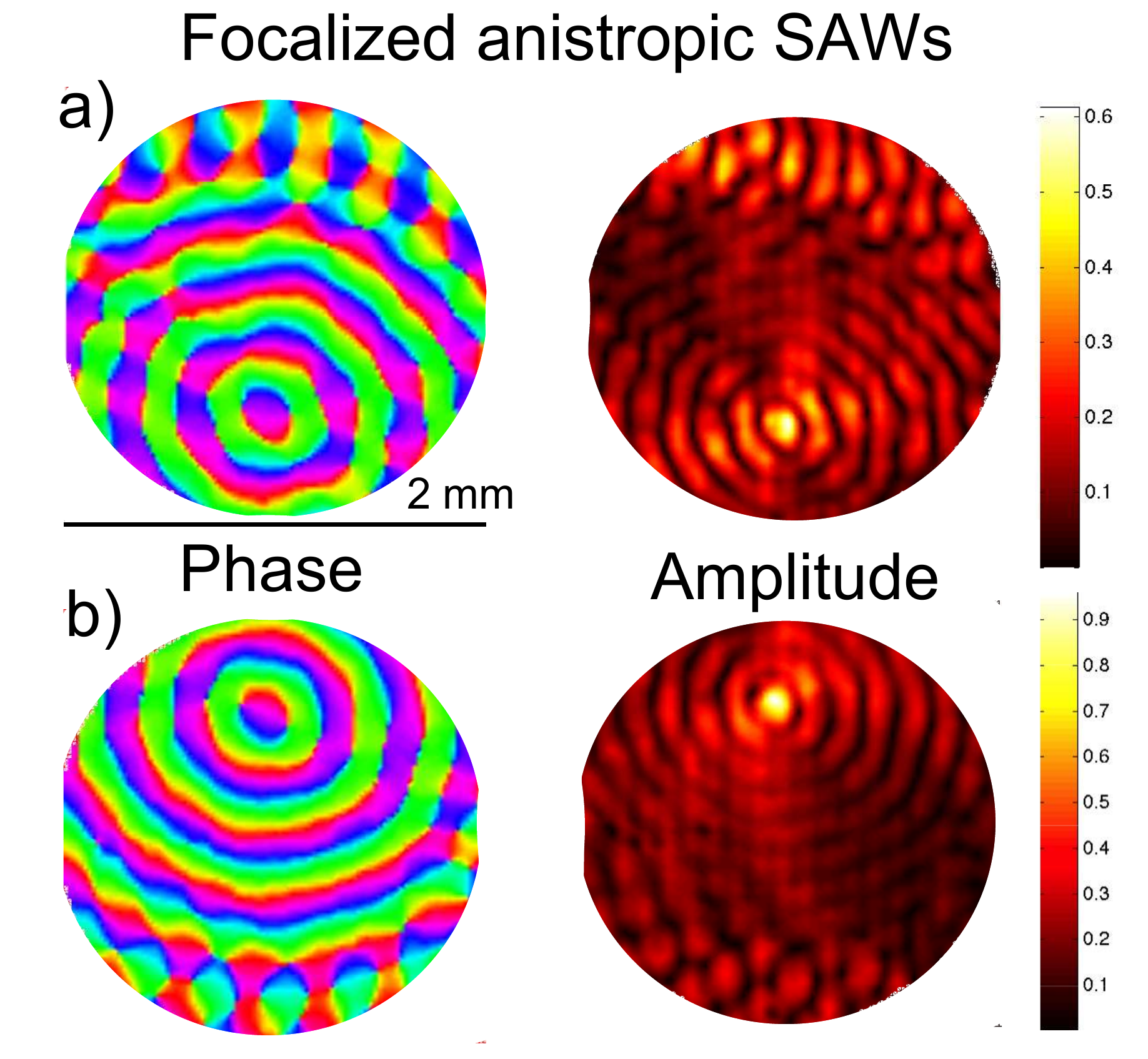}
\end{center}
\caption{Phase and amplitude of focalized propagative anisotropic SAWs converging at two different focal points (a and b). The maximum peak to peak amplitude for the two different focal points are respectively $60$ nm and $100$ nm. Colorbar: peak to peak amplitude $\times 10^{-2}$ nm.}
\label{fig:focalized}
\end{figure}

Then, we synthesized focalized waves with different positions of the focal point in the acoustical scene and apodization around a preferential direction $\phi_o$ (see Fig. \ref{fig:focalized}). It is important to note that contrarily to previous attempts to synthesize focalized waves for microfluidic applications with concentric IDTs \cite{rmp_friend_2011}, the anisotropy of the piezoelectric medium is considered here in the definition of the targeted wave-field:
\begin{eqnarray}
&&\mathcal{F}(r,\theta)= \frac{1}{2\pi} \int\limits_{-\pi}^{\pi}  h(\phi-\phi_0,\sigma)  e^{i k_r(\phi)r\cos(\phi-\theta)} d\phi, \nonumber \\
&& \mbox{ with }h(\phi-\phi_0,\sigma) = \exp\left[-\frac{\|\phi-\phi_0\|^2}{4\pi^2\sigma^2}\right], \nonumber
\end{eqnarray}
$(r, \theta)$ the polar coordinates, $k_r (\phi) = \omega / c_R(\phi)$, the radial wave vector, $c_R(\phi)$, the phase speed of R-SAWs in the $\phi$ direction (defining the anisotropy of the medium, $\sigma$ the aperture (here $\sigma^2$ = 0.2) and the symbol $\|$ denotes the shorter angular distance (which can be computed as follows $\|\phi-\phi_0\| = \arccos(\cos(\phi-\phi_0))$). The function $h$ then represents an apodization over an aperture $\sigma$.

Consideration of the anisotropy is indeed essential to ensure real focalization of the acoustic wave. Excellent results are obtained (see Fig. \ref{fig:focalized}) with maximum normal amplitude of the R-SAW of up to $100$ nm.  Naturally, small discrepancies are observed compared to the targeted wave-field  (the phase does not exactly follows the slowness curve) owing (i) to the finite numbers of transducers used for the field synthesis and (ii) to the finite number of measurement points obtained with the interferometer and used for the reconstruction of the experimental acoustic field.

\begin{figure}[htbp]
\begin{center}
\includegraphics[width=0.45\textwidth]{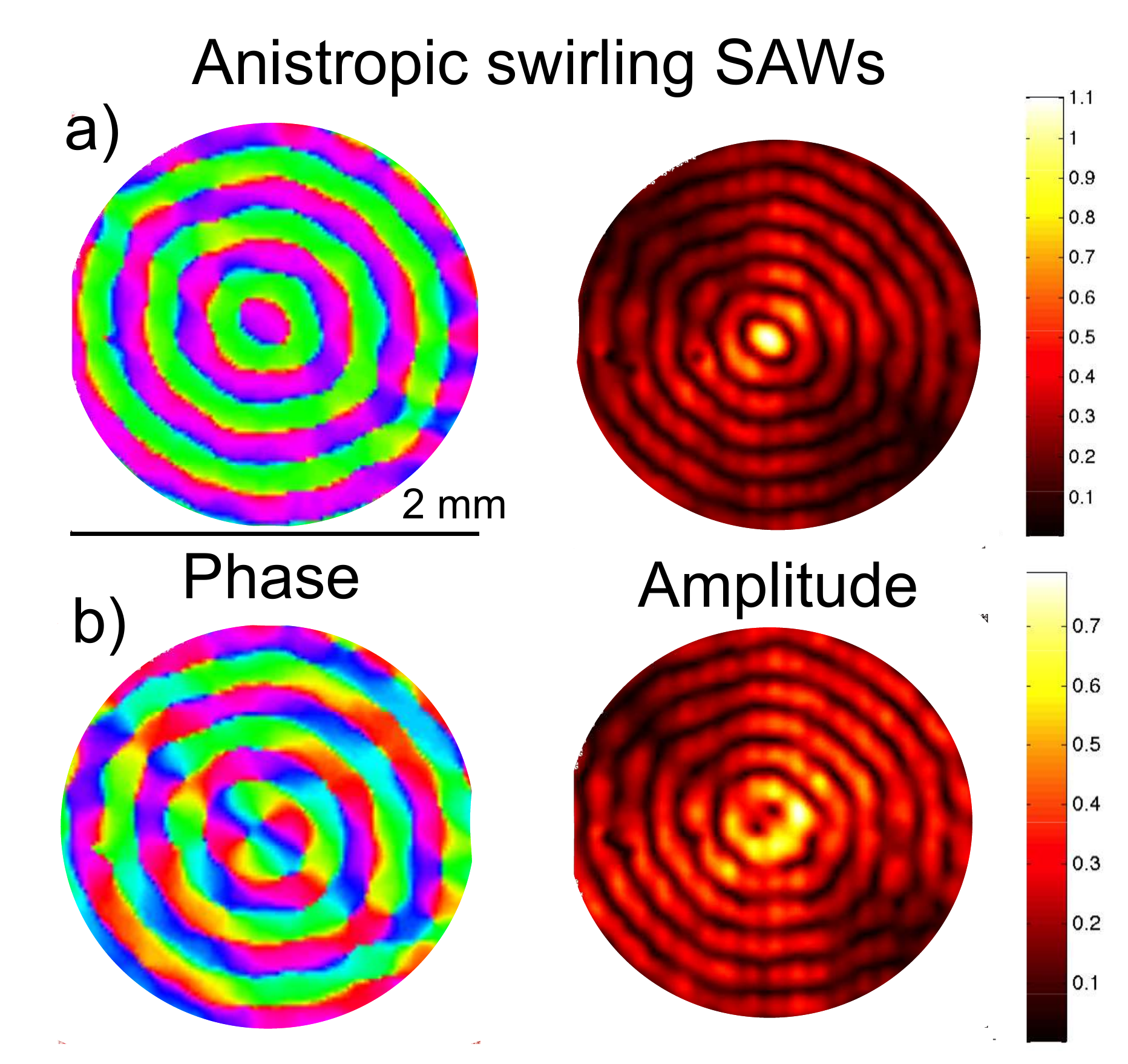}
\end{center}
\caption{Phase and amplitude of anisotropic swirling SAWs of topological order 0 and 2 (a and b). The maximum peak to peak amplitude of the $\mathcal{W}_0$ and $\mathcal{W}_2$ swirling SAWs are respectively $108$ nm and $80$ nm. Colorbar: peak to peak amplitude $\times 10^{-2}$ nm.}
\label{fig:swirling}
\end{figure}

Finally, anisotropic swirling SAWs of topological order $0$ and $2$ have been synthesized with the new transducers array presented in this paper (see Fig. \ref{fig:swirling}). The targeted wave-field is defined according to the formula introduced in ref. \cite{prap_riaud_2015}:
\begin{equation}
\mathcal{W}_l(r,\theta) = \frac{1}{2\pi i^l}\int_{-\pi}^{+\pi}  e^{i l \phi + i k_r(\phi) r \cos(\phi - \theta)} d\phi,
\label{eq: anisotropic beam}
\end{equation}
with $l$ the topological order of the swirling SAW. It is interesting to note that swirling SAWs of order $0$ are nothing but focalized waves with no angular apodization. These waves are not appropriate for particles tweezing or vorticity control since they do not have phase singularity (and thus dark spot) at their center to trap particles and they do not carry angular momentum contrarily to higher order swirling SAWs. Nevertheless, they can be of much practical use for applications where high intensity is required at the focal point.

\subsection{Droplet manipulation with the platform}

\begin{figure}[htbp]
\begin{center}
\includegraphics[width=0.45\textwidth]{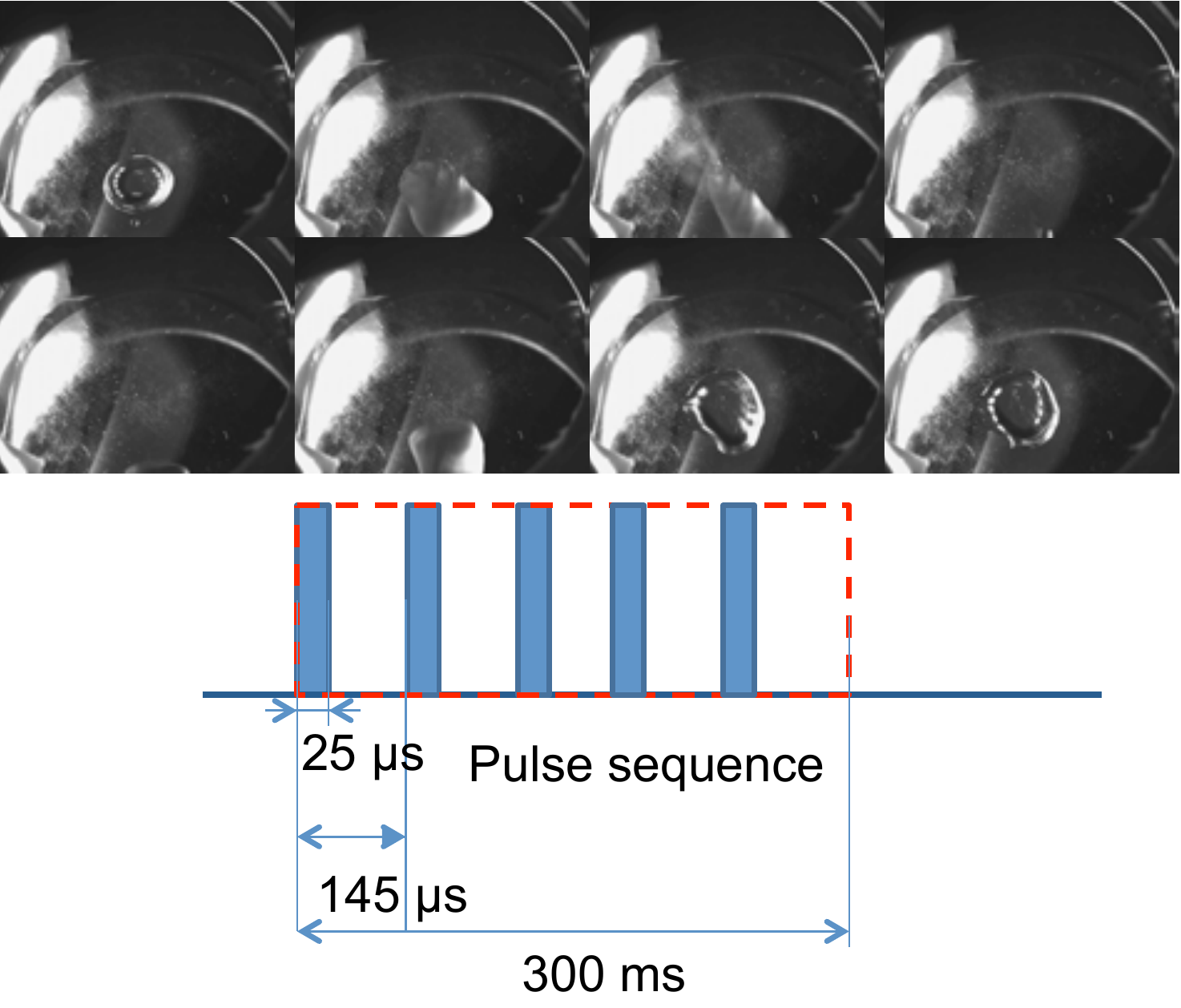}
\end{center}
\caption{Stack of images showing the displacement forward and backward of a $2 \; \mu l$ droplet in the acoustical scene. The displacement is obtained with focalized waves propagating successively in the opposite directions The time elapsed between two successive images is $53$ ms. See also movie S1.}
\label{fig:displacement}
\end{figure}

These specific wave-fields have been used to perform operations on water droplets of initial volume $2 \; \mu l$. Since it is not possible to synthesize continuous wave-fields with the programmable electronics, some burst of duration $25 \; \mu$s, carrying frequency $11.9$ MHz and variable repetition rates of a few $kHz$ have been used for droplet actuation (for each operation, the exact sequence used for actuation is described on the corresponding figure). Since droplet displacement is the result of cumulative nonlinear effects (radiation pressure and acoustic streaming), and the characteristic hydrodynamic times associated with droplets motion are slow compared to the repetition rates, the forcing is essentially seen as a continuous forcing. Nevertheless, the repetition rates of several $kHz$ are compatible with droplets high order inertio-capillary vibrations \cite{prsl_rayleigh_1879,pof_qi_2008,apl_baudoin_2012,l_blamey_2013}, which may explain why spurious atomization of the drop has been observed. Indeed, both the frequency of the excitation and the acoustical powers used in the present experiments are compatible with droplet nebulization \cite{pof_qi_2008} (larger power than usual were used in the present experiments to overcome the retention of the contact line). Nevertheless, this shortcoming could be simply overcome with appropriate hydrophobic treatment of the surface with low hysteresis \cite{nc_crudden_2014}.

\begin{figure}[htbp]
\begin{center}
\includegraphics[width=0.45\textwidth]{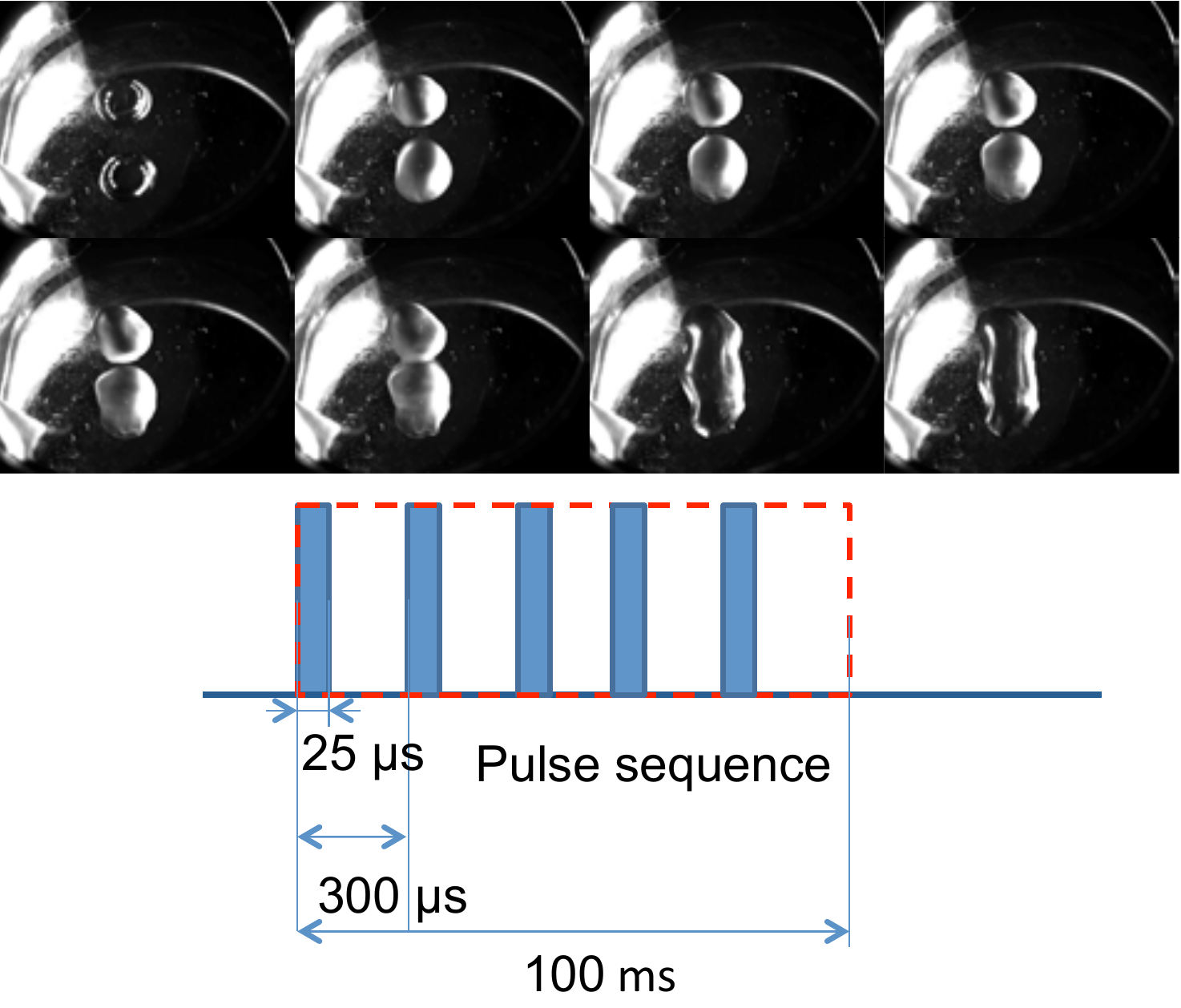}
\end{center}
\caption{Stack of images showing the fusion of two droplets of $2 \; \mu l$. The fusion is obtained with a centered swirling SAWs of topological order 2. The time elapsed between two successive images is $27$ ms. See also movie S3.}
\label{fig:fusion}
\end{figure}

\begin{figure}[htbp]
\begin{center}
\includegraphics[width=0.45\textwidth]{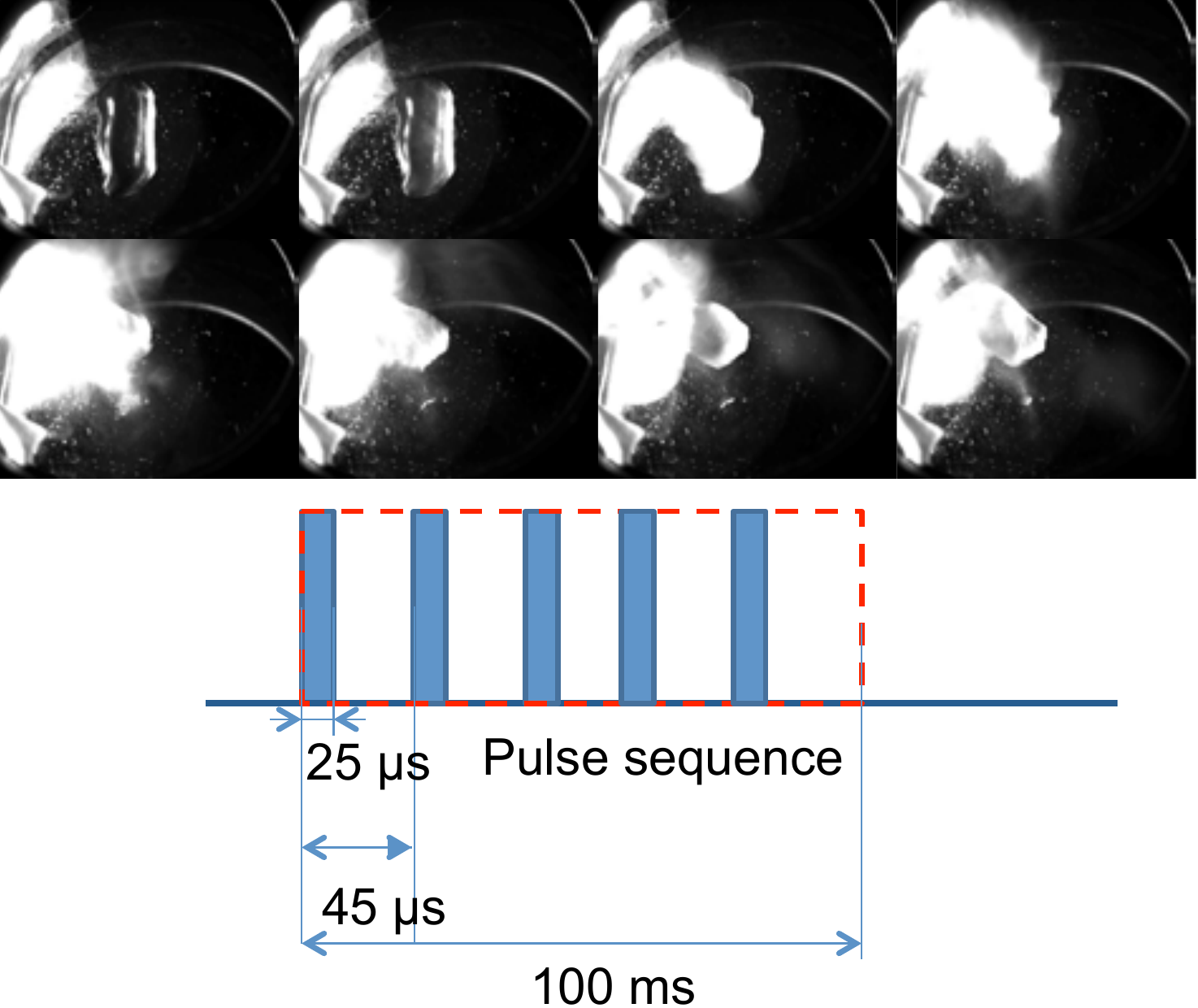}
\end{center}
\caption{Stack of images showing the atomization (nebulization) of a $2 \; \mu l$ droplet. It is obtained with a high intensity centered swirling SAW of order $0$ (annular wave). The time elapsed between two successive images is $13$ ms. See also second part of movie S3.}
\label{fig:atomization}
\end{figure}

\begin{figure}[htbp]
\begin{center}
\includegraphics[width=0.45\textwidth]{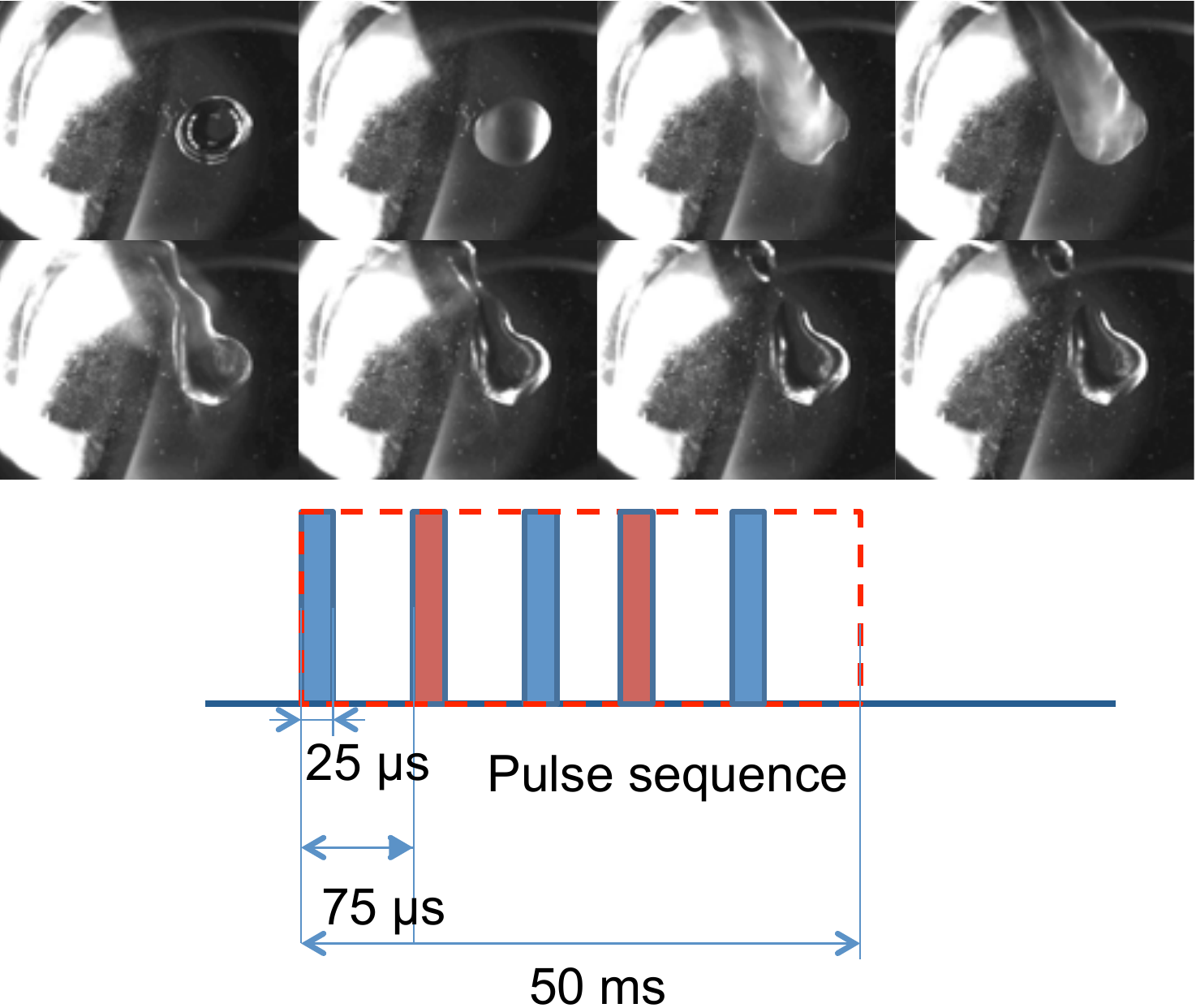}
\end{center}
\caption{Stack of images showing the asymmetric division of a $2 \; \mu l$ droplet into two daughter droplets of different volumes. It is obtained with waves focalized at two focal points situated on both sides of the drops and with angular apodization. The time elapsed between two successive images is $27$ ms. See also movie S2.}
\label{fig:division}
\end{figure}

Droplet displacement (Fig. \ref{fig:displacement}, Movie S1), fusion (Fig. \ref{fig:fusion}, movie S3) and atomization (Fig. \ref{fig:atomization}, second part of movie S3) were obtained respectively with focalized waves, swirling SAWs of second and zero order. The characteristic of the wave-field are summarized on the corresponding figure. For the droplet division (Fig. \ref{fig:division}, movie S2), we used a different method than the one proposed in ref. \cite{loc_collignon_2015}: we alternatively synthesized some burst of focalized waves with two different focal points as presented in the previous subsection. With this specific wave-field, we were able to separate droplets even at these high contact line hysteresis.

\section{Discussion}
In this paper, we have shown that the combination of interdigitated transducers array (IDTA) and the inverse filter technique enables the synthesis of the most common surface acoustic wave-fields used for actuation of fluids at the microscale:  plane waves in different directions, anisotropic focalized SAWs and anistropic swirling SAWs. As a proof of concept, we show that this SAW toolbox enables to perform many basic operations required in droplet based digital microfluidics: droplet displacement, division, fusion and atomization. Since it is virtually possible to synthesize any acoustic wave-field compatible with the anisotropy of the substrate, it might be easily demonstrated that it is possible to perform other operations previously demonstrated in the literature such as micro-mixing, collective particle manipulation with standing waves, or jetting. In the same way, it might be possible to demonstrate the versatility of this platform for fluid samples manipulations in microchannels. 

The most thrilling perspectives of this SAW toolbox nevertheless lie in the development of new operations that cannot be achieved with other techniques, such as selective 3D manipulation of one or a few particles or cells (displacement, rotation), precise vorticity control independently of the boundary conditions with swirling SAWs, or collective manipulation of several sessile droplets. This last operation is indeed not possible with a limited number of transducers since (i) there is a shadow zone behind each droplet that prevents the displacement of a second droplet situated behind it due to the strong absorption of Rayleigh waves by liquid samples, and (ii) it would be necessary to localize the droplets position in real time. This problems can be overcome with IDTA. Indeed, the combination of the acoustic waves emitted by each transducers would enable better spatial coverage of the substrate and thus the elimination of shadows zone. Moreover the IDTs can be used not only as actuators but also as sensors of the acoustical echo as demonstrated by Alzuaga et al. \cite{pieee_alzuaga_2003,ieeeuffc_bennes_2007} and Renaudin et al. \cite{sab_renaudin_2006} to determine the location of the drop without extra visualization setup.

This work opens prospects toward the design of potentially the most versatile toolbox for active control of small fluid samples, for microfluidics and biological applications.


%
\section*{Acknowledgment}

The authors would like to thank Pr. James R. Friend and Pr. David. W. Greve. for their invitation to IEEE International Ultrasonic Symposium 2015 in Taipei, that enabled our group to present these results.

\ifCLASSOPTIONcaptionsoff
  \newpage
\fi



\bibliographystyle{IEEEtran}
%



%

\begin{IEEEbiography}[{\includegraphics[width=1in,height=1.25in,clip,keepaspectratio]{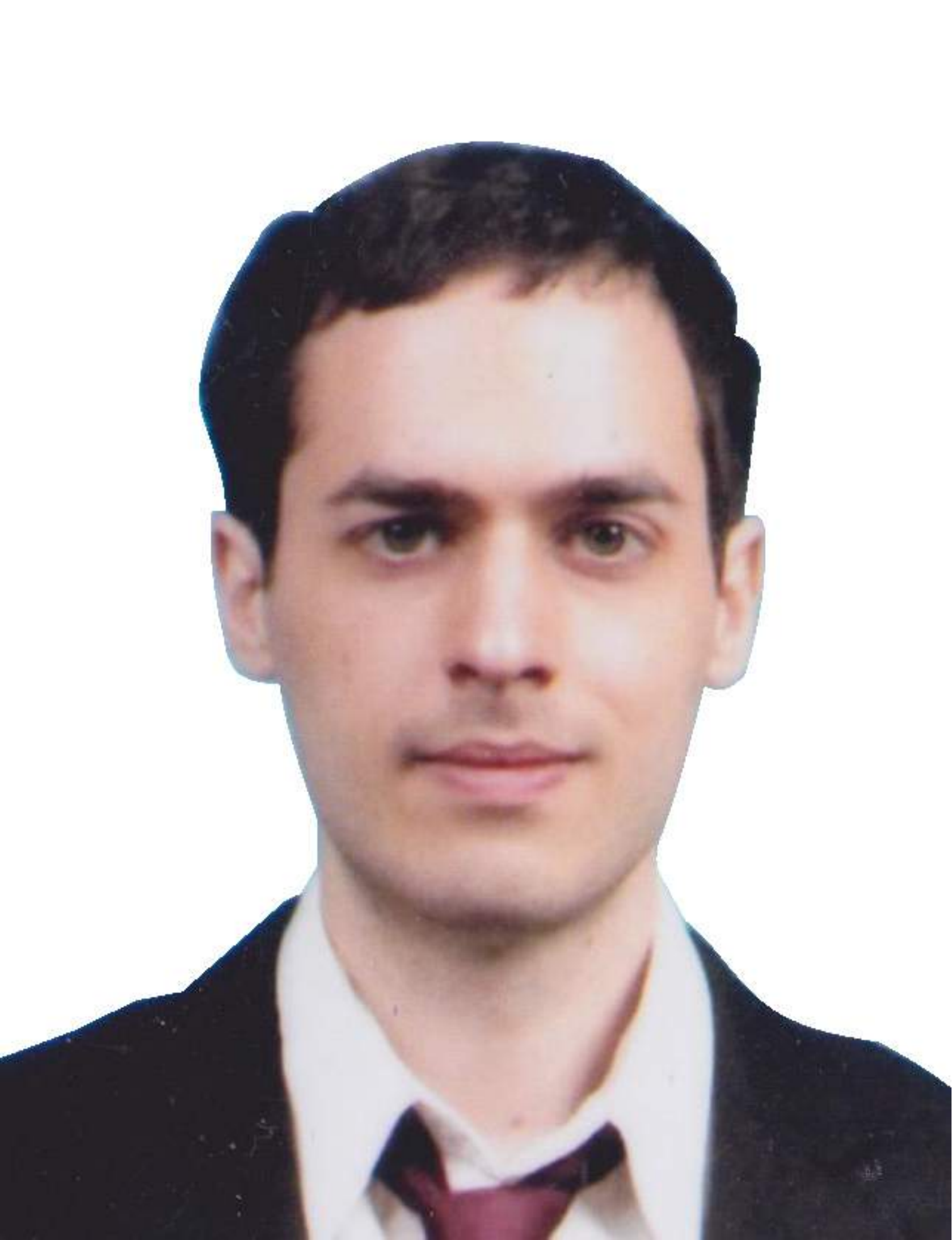}}]{Antoine Riaud}
was born in 1989 in Toulouse, France. He received his M.Sc degree in industrial and chemical engineering conjointly from Centrale Lille (France) and Tsinghua University (China). He is currently PhD candidate at Institut d'Electronique, Microelectronique et Nanotechnologies and Institut des Nanosciences de Paris.  His research interests orbit around microfluidics and include Lattice-Boltzmann simulation methods, surface acoustic waves actuation of fluids and particles and quantitative biology.
\end{IEEEbiography}

\begin{IEEEbiography}[{\includegraphics[width=1in,height=1.25in,clip,keepaspectratio]{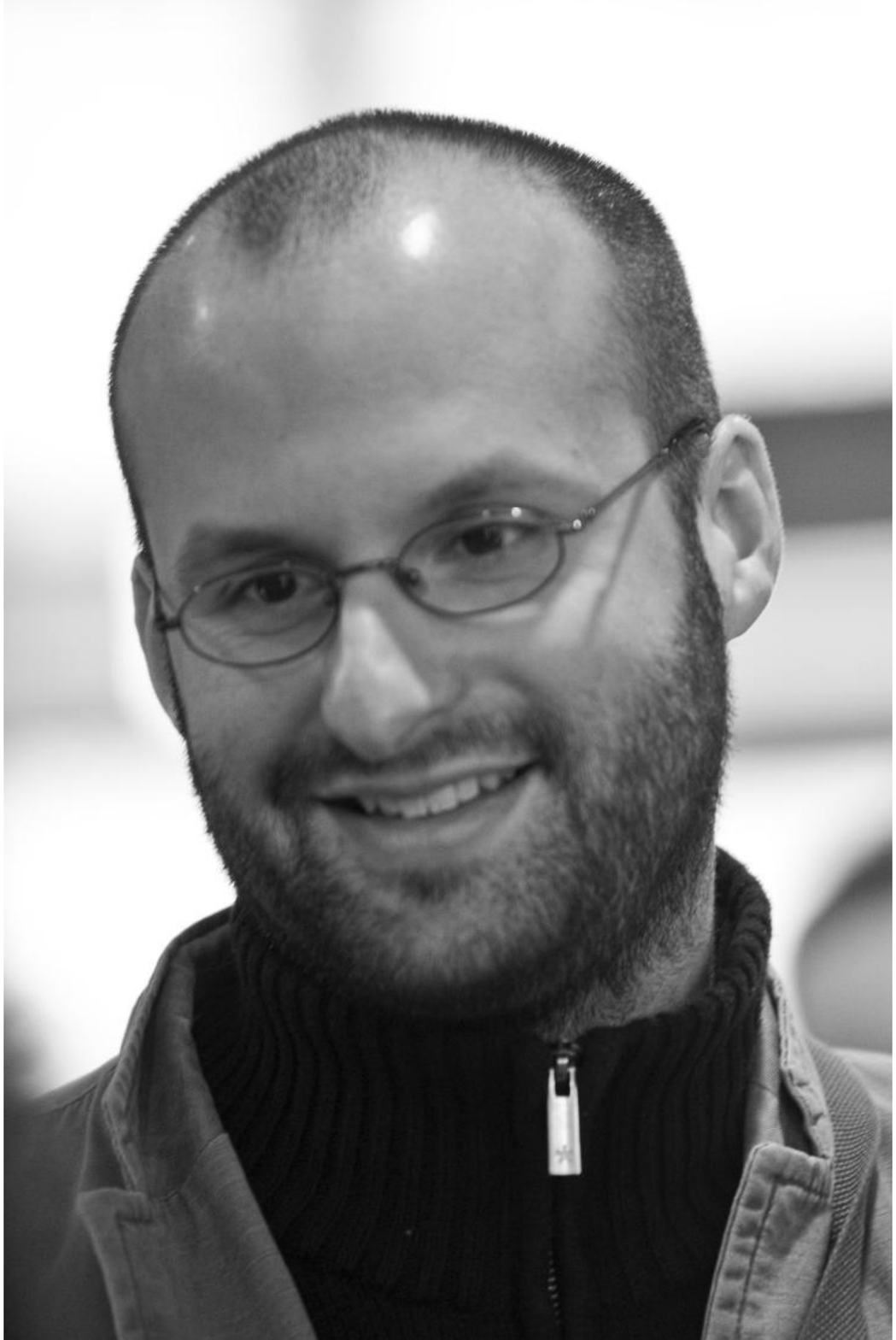}}]{Michael Baudoin}
 was born in Massy (France) in 1980. He received a M.Sc degree in mechanical engineering from the Ecole Nationale Sup\'{e}rieure de Technique Avanc\'{e}es (Paris Institute of Technology) in 2004, a M.Sc degree in fluid mechanics from Universit\'{e} Pierre et Marie Curie in 2004 and  a PhD in fluid mechanics and acoustics from Universit\'{e} Pierre et Marie Curie in 2007. In 2008, he joined LadHyX laboratory at Ecole Polytechnique for a postdoc. He is now associate professor at Universit\'{e} de Lille. His research interests lie at the interface between acoustics, microfluidics and microsystems.
\end{IEEEbiography}
%
%
\begin{IEEEbiography}[{\includegraphics[width=1in,height=1.25in,clip,keepaspectratio]{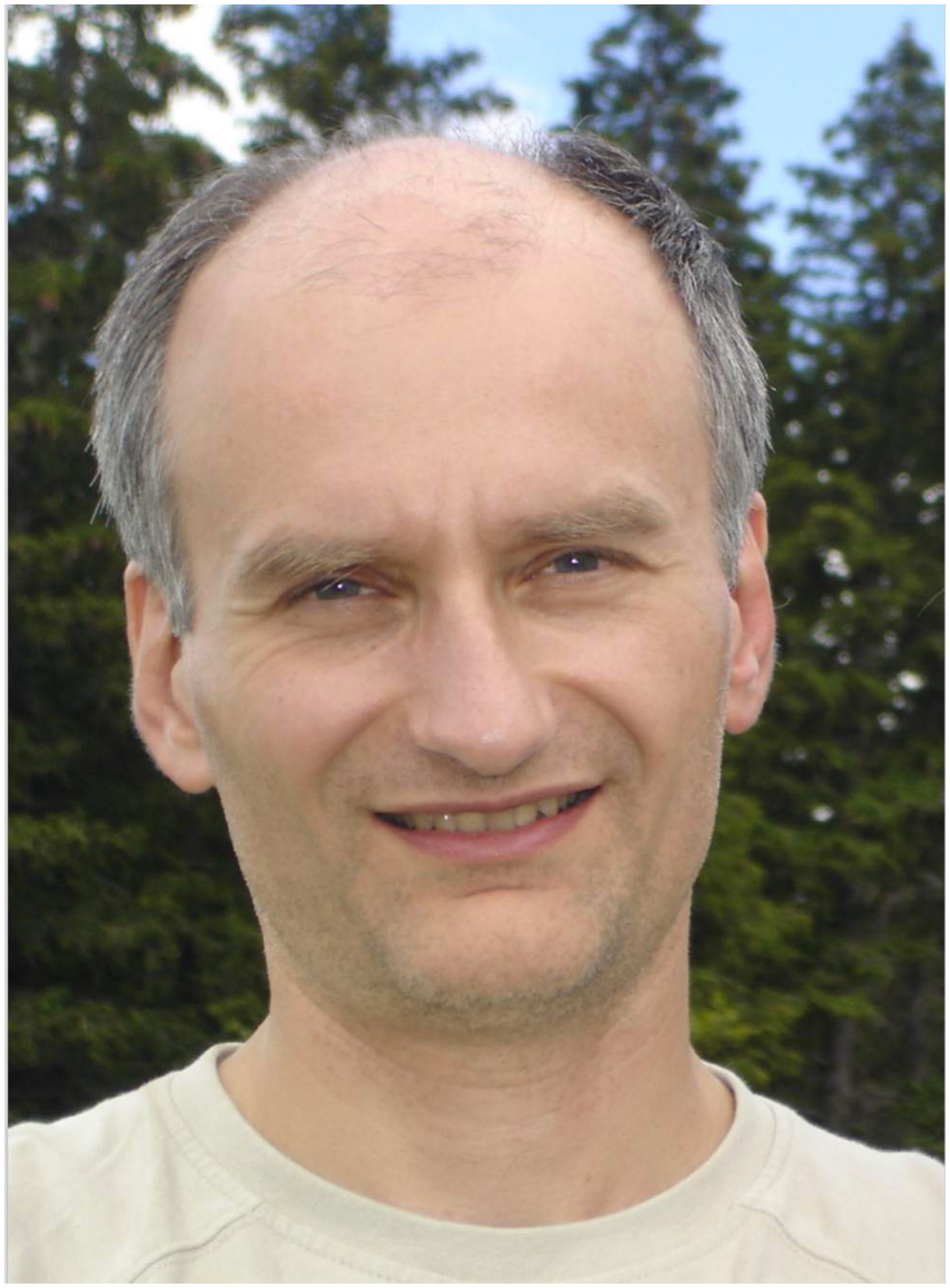}}]{Jean-Louis Thomas}
was born in 1966 in Paris, France. He received his M.Sc degree in physics (1990) from the University of Paris 6. In 1994, he received the Ph.D. degree in physics from the same university for his work on time reversal mirror. He got a position of researcher at CNRS in 1994. Since 2005, he works at « Institut des NanoSciences de Paris », CNRS and Pierre et Marie Curie university, Paris, France. His research activities include adaptative focusing in heterogeneous media, nonlinear acoustics, acoustical vortices and sonoluminescence.
\end{IEEEbiography}

\begin{IEEEbiography}[{\includegraphics[width=1in,height=1.25in,clip,keepaspectratio]{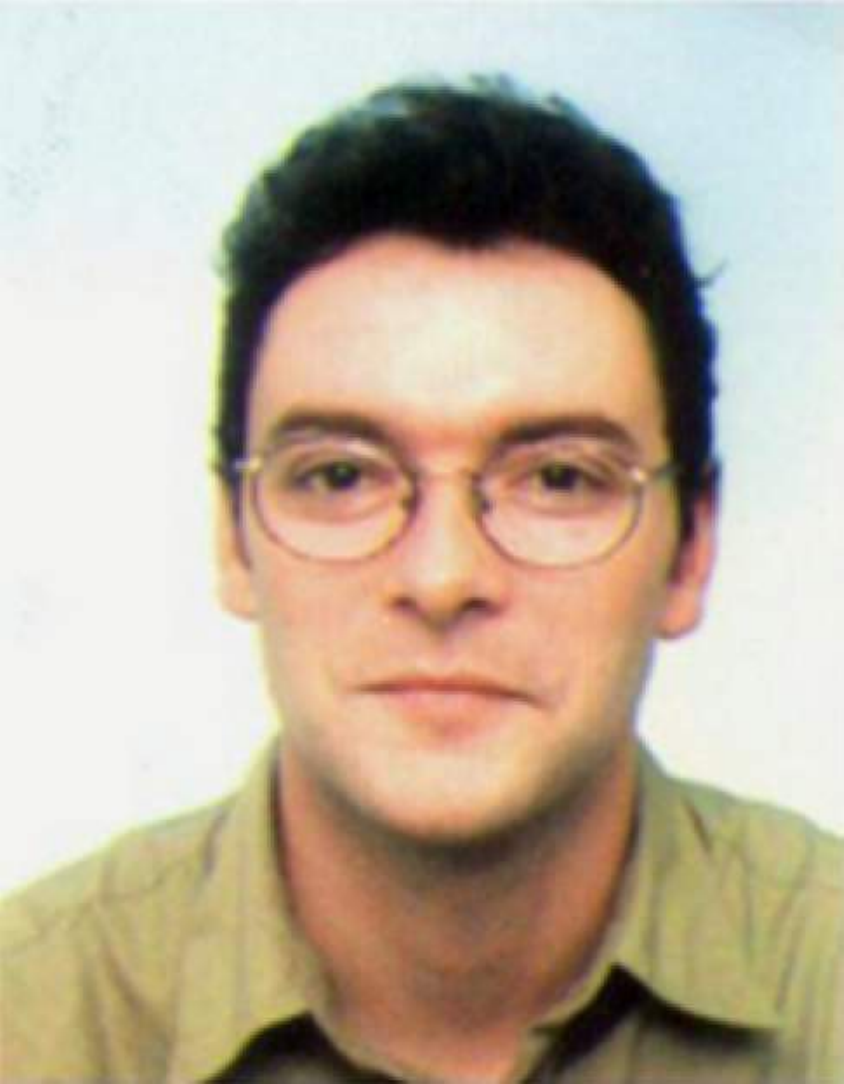}}]{Olivier Bou Matar}
was born in Villeneuve sur Lot in 1970. He received the Ph. D. degree in Acoustics from the University of Tours, France, in 1997. In 1998, he joined the Blois Technological Institute as Associate Professor of electronic and communication systems. Since 2005, he became a full Professor at Ecole Centrale de Lille and works in the Magnetoacoustics research group of the Institute of Electronics Microelectronics and Nano-technologies (IEMN, CNRS 8520), where his research focuses on nonlinear magneto-elasticity and magnetostrictive films for microsystems. He is also involved in the Joint International Franco-Russian-Ukrainian Laboratory LIA LICS. He has gained a great experience in numerical simulation of nonlinear acoustic wave propagation, in the development of ultrasonic imaging systems ("Non-linear imaging by Magneto-Acoustic Phase Conjugation (MAPC)"), and in multiphysics applications involving wave propagation in solids and liquids. He is now the head of Acoustical Department of the Institute of Electronics Microelectronics and Nanotechnologies. He has co-authored over 120 international publications in journals and proceedings mainly in the domain of nonlinear acoustic, phononic crystals and magnetoacoustic.
\end{IEEEbiography}




\end{document}